\documentclass[3p,times,procedia]{elsarticle}
\usepackage{nupha_ecrc}
\usepackage{color}

\volume{00}

\firstpage{1}

\journalname{Nuclear Physics A}

\runauth{}

\jid{nupha}

\jnltitlelogo{Nuclear Physics A}

\usepackage{amssymb}

\usepackage[figuresright]{rotating}

\begin{document}

\begin{frontmatter}



\dochead{}

\title{The sensitivity of $R_{pA}$ to color recombination effects}


\author{Korinna Christina Zapp$^{1,2}$, Guilherme Milhano$^{1,2}$ and Urs Achim Wiedemann$^{1}$}

\address{$^1$ Physics Department, Theory Unit, CERN, CH-1211 Gen\`eve 23, Switzerland\\
$^2$ CENTRA, Instituto Superior T\'ecnico, Universidade de Lisboa, Av. Rovisco Pais, P-1049-001 Lisboa, Portugal}

\begin{abstract}
In hadronization models with color recombination, partons are allowed to regroup into color singlet structures that are different from those determined by the
perturbative parton shower. This aims at modeling the possibility that soft interactions of partons with the underlying event can change color connections. 
If such an effect is at play in proton-proton collisions, it may be expected to be enhanced in proton-nucleus collisions due to the higher color charge density in 
the underlying event. Here, we provide a qualitative argument that color recombination effects could lead to a multiplicity dependent hardening 
of single inclusive hadron spectra that dies out very weakly ($\propto 1/p_\perp$) with increasing transverse momentum. We present results of a (conservative) 
model implementation in the cluster hadronization model of the SHERPA event generator. In this model, we find that color recombination effects harden indeed the
single inclusive hadron spectra without affecting the jet spectra, but that this effect does not depend significantly on underlying event activity. We explain this
model feature and we argue why, in general, data on proton-nucleus collisions can help to constrain hadronization models used in proton-proton event generators. 
\end{abstract}


\end{frontmatter}
\vspace{0.5cm}



In proton-lead (pPb) collisions at $\sqrt{s_{\rm NN}} = 5.02$ TeV, the CMS and ATLAS collaborations have reported a charged hadron nuclear modification
factor $R_{pPb}^{h^\pm}(p_\perp)$ that exceeds unity by 20 - 50 \%  at the highest transverse momenta analyzed so far 
($ 20\, {\rm GeV} < p_\perp < 100\, {\rm GeV}$)~\cite{Khachatryan:2015xaa,ATLAS:2014cza}.
In marked contrast, the nuclear modification factor $R_{pPb}^{\rm jet}(E_\perp)$ of jets is compatible with unity for $E_\perp > 50$ GeV\cite{ATLAS:2014cpa,CMS:2014qca,Adam:2015hoa}.
Also, jet fragmentation functions in pPb show only mild deviations from those measured in pp collisions~\cite{CMS:2015bfa,ATLAS:2015mla}, 
and effects from the nuclear modification of parton distribution functions~\cite{Helenius:2015wda} are much smaller than the reported 20 - 50 \% excess at high $p_\perp$.
These data seem to point to a hadronization mechanism that is harder in pPb than in pp. As we explain below, color recombination effects may account for such an effect since they can
lead to a multiplicity-dependent hardening of single inclusive hadron spectra without affecting jet spectra. This was the first motivation for the present study.
However, we anticipate already here that our study does not identify an effect that could account quantitatively for the observed discrepancy between $R_{pPb}^{h^\pm}(p_\perp)$ and $R_{pPb}^{\rm jet}(E_\perp)$. 
We also caution that the CMS collaboration presented recently an updated physics analysis summary that does not exclude a substantial reassessment of the baseline entering their
determination of $R_{pPb}^{h^\pm}(p_\perp)$ (see Fig.15 of Ref.~\cite{CMS:2015bfa}). So, it is currently unclear whether a theory explanation of this discrepancy is needed. 

\begin{figure}
\includegraphics[width=0.7\linewidth]{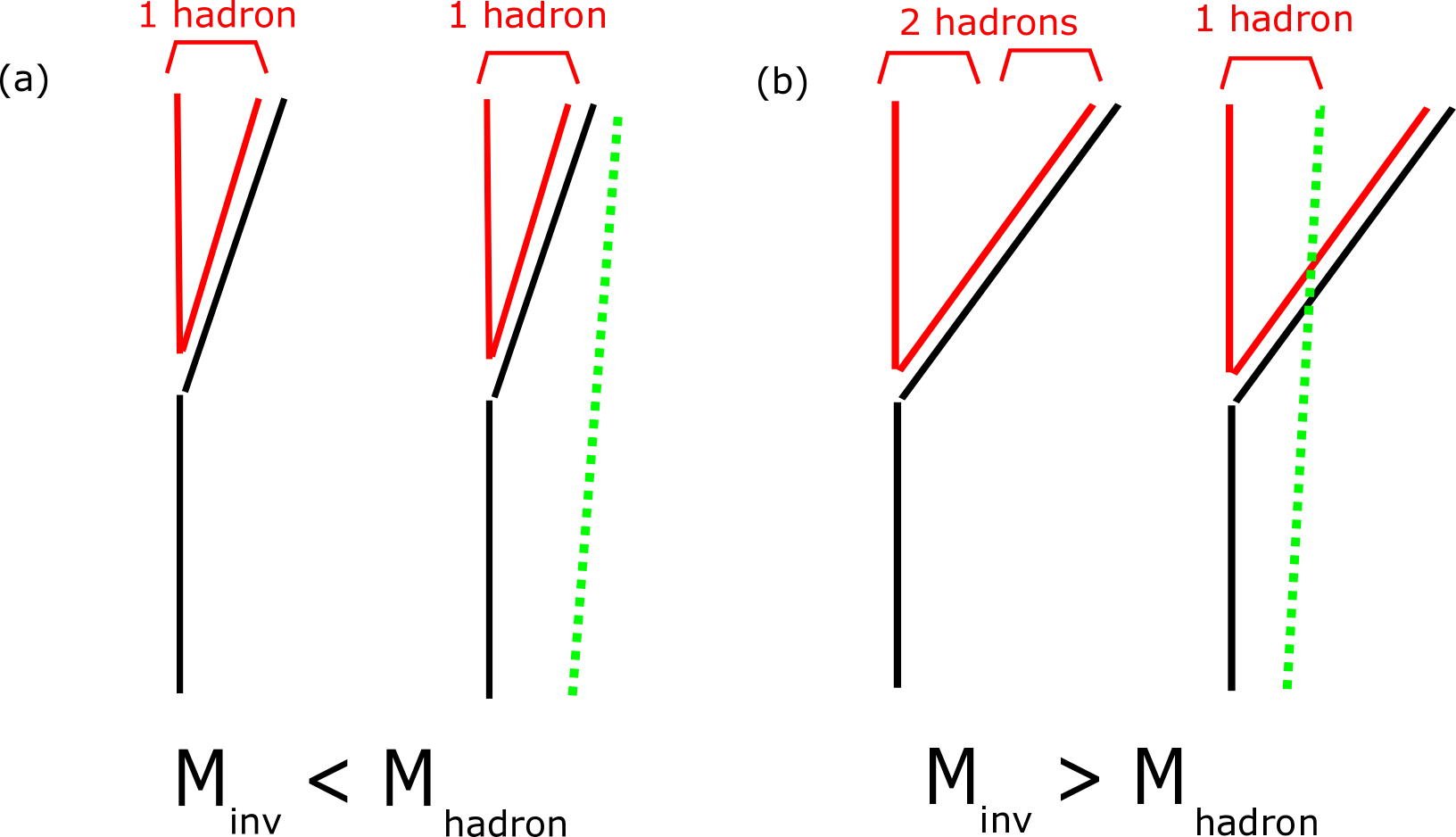}
\caption{
Schematic picture of a simplified cluster hadronization mechanism. Color recombination effects recombine soft partons of the underlying event (green lines) if they are 
sufficiently close in phase space to high-$p_\perp$ partons. This figure is discussed in more detail in the text.}
\label{fig1}
\end{figure}

Let us consider the simulation of a hadronic collision up to the stage when 
all partons have been evolved to some hadronic scale $Q_0$. The partonic distribution of the entire event can then be grouped into a set of color singlet clusters of known 
invariant masses $M^2$. Most generally, a cluster hadronization model will treat these color singlets as resonance-like states that decay into hadronic distributions according to a 
model-dependent but process-independent prescription. Since partons of the highest transverse momentum, say $p_{\perp}$, are typically the hardest component of a 
parton shower, they are color connected to partons of subleading energy $k_\perp$ in this shower, and the difference between these partons in momentum space sets
the contribution to the invariant mass in the color singlet,
\begin{equation}
	M_{\rm inv}^2 = p_\perp\, k_\perp\, \left[ \cosh\left(\eta_p-\eta_k \right) - \cos \left(\phi_p-\phi_k \right) \right] \approx p_\perp\, k_\perp\,R^2\, ,
         \label{eq1}
\end{equation}
where $\eta$ and $\phi$ denote the pseudorapidity and the azimuthal angle of the parton's momentum, and $R^2 =  \left(\eta_p-\eta_k \right)^2 + \left(\phi_p-\phi_k \right)^2$
is the standard squared distance between the two partons in the $\Delta\eta-\Delta\phi$-plane. For clusters of large transverse momentum, one typically obtains 
in this way an $M_{\rm inv}^2$-distribution that is such that some fraction of the clusters are of sufficiently low invariant mass
to be mapped directly to a hadron of high transverse momentum (Fig.~\ref{fig1}a), while the other fraction of hadrons has a too large invariant mass and will
undergo resonance-like decays into at least two hadrons (Fig.~\ref{fig1}b). The high-$p_\perp$ hadron spectrum is dominated by hadrons from clusters of low 
invariant mass, since any hadronic decay will further degrade the transverse momentum of the leading hadron.
%

 Color recombination mechanisms assume that partons of transverse momentum $p_\perp$ can form a color singlet cluster with an arbitrary parton of low-$k_\perp$ (green dashed line in Fig.~\ref{fig1}) if 
their invariant mass (\ref{eq1}) is minimal. In pp or pA collisions, the underlying event activity is low and less than one parton is found in a typical opening cone $R$.
The probability of a perturbative color singlet to recombine with a soft component of the underlying event is then much smaller for a singlet of low invariant mass, since the corresponding $R$ is smaller. 
On the other hand, since color reconnection reduces the invariant mass of the leading cluster, it decreases the probability of further resonance-like decays of this cluster (Fig.~\ref{fig1}b); this effect
makes the resulting hadron spectrum harder. And since the probability of color reconnection increases proportional to the density of color components in the underlying event, this effect will be 
enhanced in pPb compared to pp. {\it We thus expect that a color recombination mechanism invoked in pp-collisions will be enhanced by the ratio of the typical minimum bias multiplicities,
i.e. by a factor $\approx 3$, in pPb collisions.}

Colour reconnection or rearrangement effects have been discussed in a variety 
of contexts over the last decades. It was noted 
in~\cite{Sjostrand:1987su} that a minimisation of the length of colour strings 
is required to describe the increase of the average transverse momentum 
$\langle p_\perp \rangle$ with multiplicity in minimum bias events at hadron 
colliders.
A large multiplicity is reached by having more multiple parton interactions 
(MPI) rather than an increase in hardness of produced jets.  Here, two effects have 
to be distinguished, namely the initial assignment of colour flow to the 
individual MPI and a rearrangement of colour in the final state prior to or 
during hadronisation (on which our discussion focusses). Both cannot be 
derived from first principles but have to be modelled. Modern Monte Carlo event 
generators~\cite{Bahr:2008pv,Sjostrand:2014zea,Gleisberg:2008ta} follow a 
principle of minimising string length when assigning colour flow to MPI and 
include at least one colour reconnection 
model~\cite{Gieseke:2012ft,Argyropoulos:2014zoa,Winter:2003tt}. 

SHERPA has a cluster hadronisation model~\cite{Winter:2003tt,Gleisberg:2008ta}, the details of which are not relevant for 
this study. After the perturbative stage (matrix elements, MPI and parton 
showers) the produced partons (i.e.\ (anti)quarks and gluons) are arranged 
into colour ordered singlets. Then, the gluons are split into quark-antiquark 
or antidiquark-diquark pairs. Pairs of matching colour form the first 
generation of clusters retaining the same ordering as before. This means that 
neighbouring clusters each contain one (anti)quark coming from the splitting 
of the same gluon. This ordering plays a role in the colour reconnection 
phase, which follows next.

During colour reconnections, pairs of clusters $(12)$ and $(34)$ are allowed to 
swap colour and thus re-arrange themselves. The probability for this to happen 
has the form
\begin{equation}
\label{Eq::reconnprob}
P_{\rm swap} = \mathcal{P}_{\rm cr} \cdot \frac{w_{1423}}
{w_{1234}+w_{1423}}\cdot
\left( \frac{1}{N_{\rm c}^2} \right)^d\, ,\qquad
w_{ijkl} = \frac{t_0}{t_0 + 4m_{ij}m_{kl}}\ ,
\end{equation}
where $m_{ij}$ is the invariant mass of the pair consisting of partons $i$ and 
$j$, $d$ is the number of clusters between the two clusters under 
consideration in the colour ordered list, $t_0$ is $O(1
{GeV^2})$ and ${\cal P}_{\rm cr}$ is a parameter regulating the strength 
of colour reconnections. The kinematic factor in (\ref{Eq::reconnprob}) obviously favours reconnections 
that decrease the clusters' invariant mass.
Clusters sharing a common ancestor gluon thus can 
reconnect without colour suppression while reconnections among clusters that 
are further apart in the original colour sequence are increasingly suppressed. 
SHERPA's model implementation is conservative in that only 
reconnections within the same (original) colour singlet system are possible.
In other colour reconnection models, larger effects are conceivable, but we have not
explored other models in the present study.  

\begin{figure}
	{\includegraphics[width=0.45\textwidth]{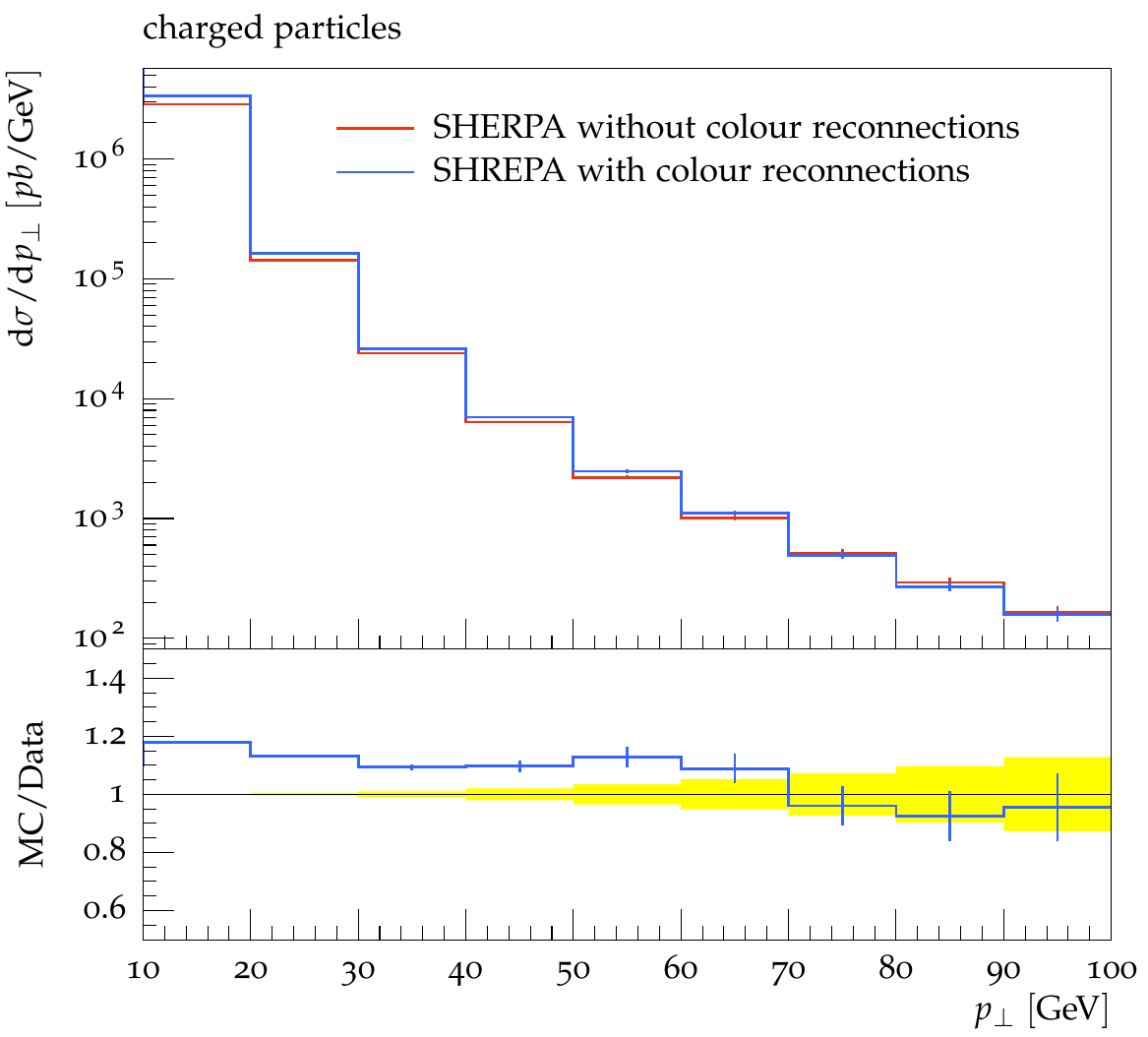} \hfill
	\includegraphics[width=0.45\textwidth]{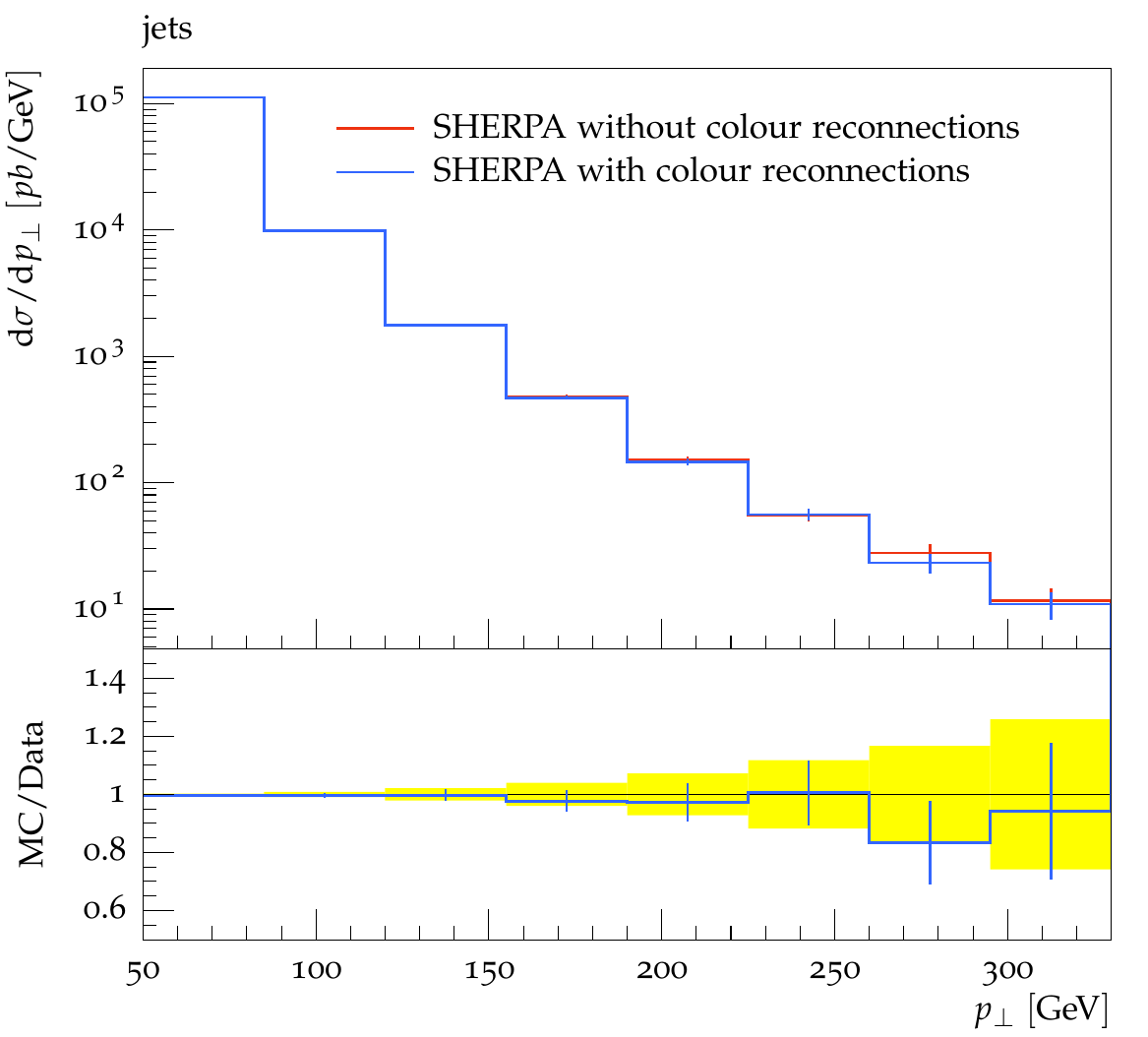}
	\caption{{\bf Left:} Charged particle $p_\perp$ spectrum with and without colour reconnections in the setting with increased underlying event activity. {\bf Right}: The same for jets (anti-$k_\perp$, $R=0.4$).}}
		\label{fig2}
\end{figure}

To simulate hadron spectra and jets with SHERPA\footnote{We used a developer's version that is for the purpose of this study similar to SHERPA\,2.1.1, with the CT10 tune and colour reconnections enabled (they are by default switched off). ${\cal P}_{\rm cr}$ is set to 0.25, which was found to be a reasonable value in earlier tuning efforts.} not only for pp but also for pPb collisions, we mimic pPb collisions by increasing the underlying event activity in pp collisions by about a factor or 3. All other settings are kept the same, including the pdf set (we do not use a nuclear pdf set, as we want to isolate the effect coming from colour reconnections).

\begin{figure}
	\centering
	\includegraphics[width=0.45\textwidth]{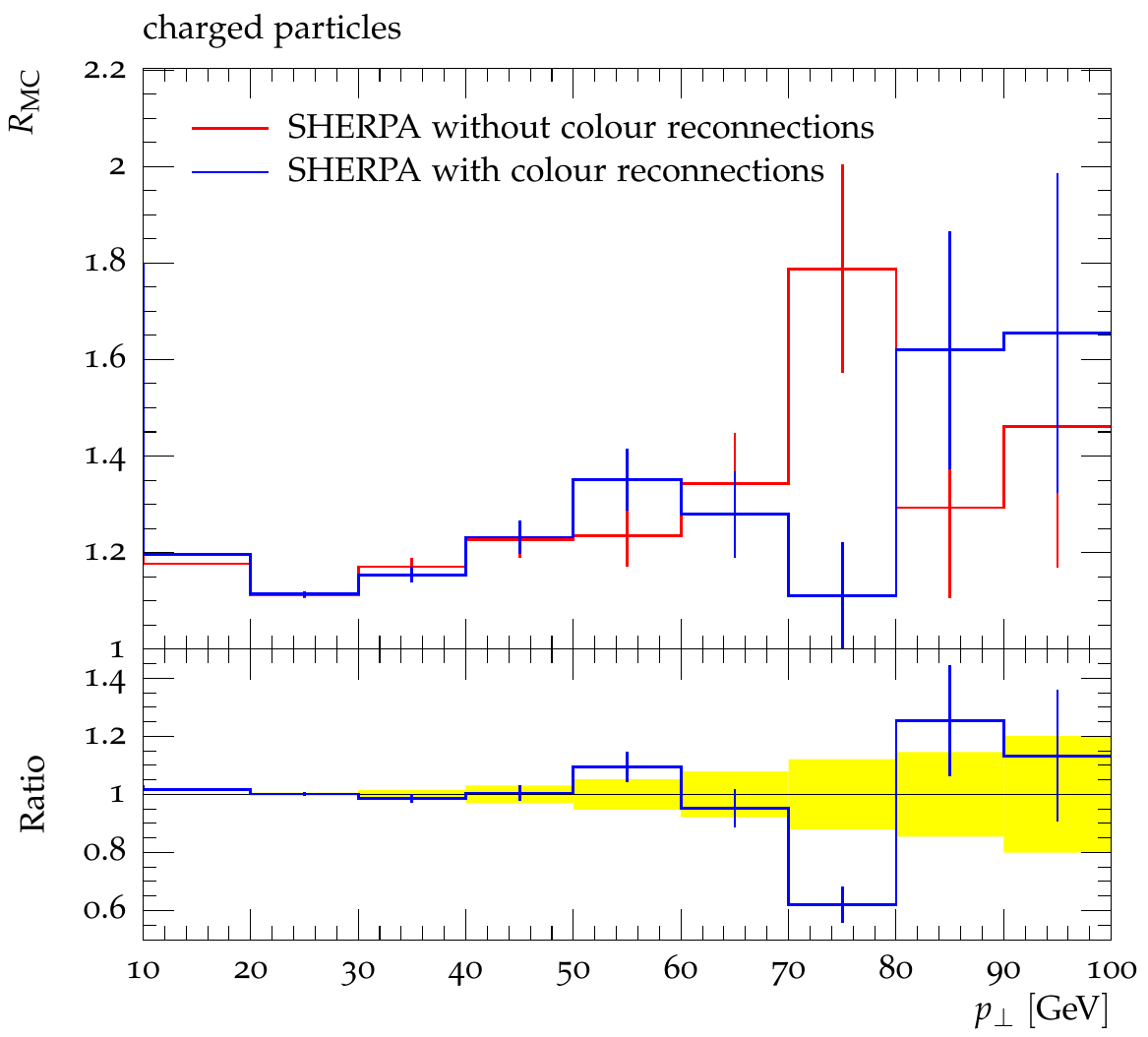}
	\caption{Charged particle $p_\perp$ spectra with increased underlying event divided by those with normal underlying event.}
	\label{fig3}
\end{figure}

Fig.~ 2 shows the charged particle and jet $p_\perp$ spectra in the set-up with increased underlying event activity with and without colour reconnections. The hadron spectrum is indeed harder with colour reconnections while the jet spectrum in unaffected. This confirms the previous argumentation. 

In Fig.~ 3, we show the ratio $R_\mathrm{MC}$ of the spectrum obtained with increased underlying event activity divided by that with normal underlying event activity. Since we mimic pPb collisions by an increased underlying event  activity of pp collisions, this ratio is a proxy for the nuclear modification factor $R_{pPb}$. One sees that this ratio is enhanced by approximately the same factor irrespective of whether colour reconnections are enabled. Therefore, the present model implementation cannot provide a dynamical explanation for the observed discrepancy between $R_{pPb}^{h^\pm}(p_\perp)$ and $R_{pPb}^{\rm jet}(E_\perp)$. 
We understand these results as follows: The enhancement of $R_\mathrm{MC}$ above unity results from event-activity dependent differences in the initial colour correlations on matrix element level that we 
mentioned earlier. On the other hand, the insensitivity of the colour recombination mechanism to the underlying event activity is likely to be a specific feature of the SHERPA implementation. Namely, since colour reconnection in SHERPA is allowed only within the same original colour singlet system, this model implementation strongly limits the probability of recombining with other partons in the underlying event. 
Therefore, the specific color recombination model implemented in SHERPA does not realize our expectation that effects of color recombination should increase with underlying event activity. 
It remains still conceivable that other less conservative color recombination effects show a different behavior.






\bibliographystyle{elsarticle-num}



\end{document}